\documentclass[10pt,draft]{dis03}
\usepackage{epsf,amsmath}

\newcommand{\NP}[1]{{\it Nucl.\ Phys.}\ {\bf #1}}

\newcommand{\PL}[1]{{\it Phys.\ Lett.}\ {\bf #1}}
\newcommand{\PR}[1]{{\it Phys.\ Rev.}\ {\bf #1}}

\newcommand{\IJMP}[1]{{\it Int.\ J.\ Mod.\ Phys.}\ {\bf #1}}
\newcommand{\be}{\begin{equation}}
\newcommand{\ee}{\end{equation}}
\newcommand{\bea}{\begin{eqnarray}}
\newcommand{\eea}{\end{eqnarray}}
\newcommand{\bfk}{\mbox{$\mathbf {k}$}}
\newcommand{\bfq}{\mbox{$\mathbf {q}$}}
\newcommand{\pup}{p^\uparrow}

\newcommand{\bfS}{\mbox{$\mathbf S$}} 
\newcommand{\Lup}{\Lambda^\uparrow} 
\newcommand{\Aup}{A^\uparrow}

\begin{document}

\title{An improved analysis of $\bfk_\perp$ effects 
for SSA in hadronic collisions\footnote{Talk delivered by U. D'Alesio
  at  the ``XI International Workshop on Deep Inelastic Scattering''
  (DIS2003), 23-27 April, 2003, St. Petersburg.} }

\author{U. D'ALESIO  and F. MURGIA \\
Istituto Nazionale di Fisica Nucleare, Sezione di Cagliari, and\\
           Dipartimento di Fisica, Universit\`a di Cagliari \\
           C.P. 170, 09042 Monserrato (CA), Italy\\
}

\maketitle

\begin{abstract}
\noindent 
The role of intrinsic transverse momentum both in unpolarized and
polarized processes is discussed.
We consider inclusive cross sections for pion production in
hadronic collisions and for Drell-Yan processes.  
We reanalyze transverse single spin asymmetries (SSA) 
in inclusive pion production, $\pup \, p \to \pi \, X$, in terms of
Sivers effect and show its
contribution to $A_{UL}$ in semi-inclusive DIS (SIDIS).
\end{abstract}

In the last years a lot of experimental and theoretical activity has
been devoted to the study of transverse single spin asymmetries (SSA)
in hadronic collisions and in semi-inclusive DIS.
In fact, perturbative QCD (pQCD) with ordinary
collinear partonic kinematics leads to negligible values
for these asymmetries.
On the other hand several experimental results
seem to contradict this expectation: 
$i)$ the large polarization of $\Lambda$'s (and other 
hyperons) produced in $p \, N \to \Lup \, X$; 
$ii)$ the large asymmetry 
observed in $\pup \, p \to \pi \, X$ and  
$\bar p^\uparrow \, p \to \pi \, X$; 
$iii)$ 
the similar azimuthal asymmetry observed in  
$\ell \, \pup \to \ell \, \pi \, X$. 

A possible way out from this situation comes from
extending the collinear pQCD formalism with the inclusion of spin and
partonic intrinsic transverse momentum, $\bfk_\perp$, effects.
This leads to
the introduction of  new spin and $\bfk_\perp$ dependent
partonic distribution (PDF) and fragmentation (FF) functions \cite{anse}.

The role of $\bfk_\perp$ effects in inclusive hadronic reactions
has been extensively studied also in the calculation of
unpolarized cross sections, where  
it has been shown that, particularly at moderate $p_{_T}$,  
these effects can be relevant \cite{ww}. 

Based on these considerations, in this contribution we present preliminary
results both for 
polarized and unpolarized cross sections (and SSA) for
inclusive particle production in hadronic collisions and SIDIS, 
using LO pQCD with
the inclusion of intrinsic transverse momentum effects.
Our main goal is 
to show
that in our LO approach unpolarized cross sections 
can be  reproduced up to an overall factor
of 2-3, compatible with NLO corrections and scale dependences,
which reasonably cancel out in SSA. 

In this approach the
unpolarized cross section for the inclusive process $A\,B\to C\,X$
reads 
\bea
d\sigma &\propto & \sum_{a,b,c,d} \hat f_{a/A}(x_a,\bfk_{\perp\,a}) \otimes
\hat f_{b/B}(x_b, \bfk_{\perp\,b}) \nonumber \\
&& \otimes\, d\hat\sigma^{ab \to c d}(x_a, x_b, \bfk_{\perp\,a},
 \bfk_{\perp\,b}) \otimes \hat D_{C/c}(z, \bfk_{\perp\,C})\,,
\label{ucr}
\eea
\noindent with obvious notations.
A similar expression holds for the numerator of a transverse
SSA ($\propto
d\,\Delta^{\!N}\!\sigma/d\sigma$), substituting, for the polarized particle
involved, the corresponding unpolarized PDF (or FF)
with the appropriate polarized one, $\Delta^{\!N}\!f$ (or
$\Delta^{\!N}\!D$).

Let us start considering the role played by the intrinsic
$\bfk_{\perp}$ in the unpolarized cross sections. 
The PDF and FF in Eq. (\ref{ucr}) are given in a
simple factorized form, and the $\bfk_\perp$ dependent part
is usually taken to have a Gaussian shape:
\begin{equation}
\hat f_{a/A}(x,\bfk_{\perp a}) = f_{a/A}(x)\,\frac{\beta^2}{\pi}\>
e^{-\beta^2\,k_{\perp a}^{\,2}}\>;
\;
\hat D_q^h(z,\bfk_{\perp h}) = D_q^h(z)\,\frac{\beta'^2}{\pi}\>
e^{-\beta'^2\,k_{\perp h}^{\,2}}\,,
\label{gk}
\end{equation}
\noindent where the parameter $\beta$ ($\beta'$) is related to the
average partonic (hadronic) $k_{\perp}$. 

Analyzing a large sample of available data for several hadronic processes and 
in different kinematical situations, 
we find that an overall acceptable account of the corresponding
unpolarized cross sections is possible by choosing, depending on the 
kinematical situation considered, 
$\beta=1.0-1.25$ (GeV/$c)^{-1}$ (that is,
$\langle\,k_\perp^2\,\rangle^{1/2}=0.8-1.0$ GeV/$c$).
\begin{figure}[!thb]
\vspace*{5.5cm}
\begin{center}
\includegraphics{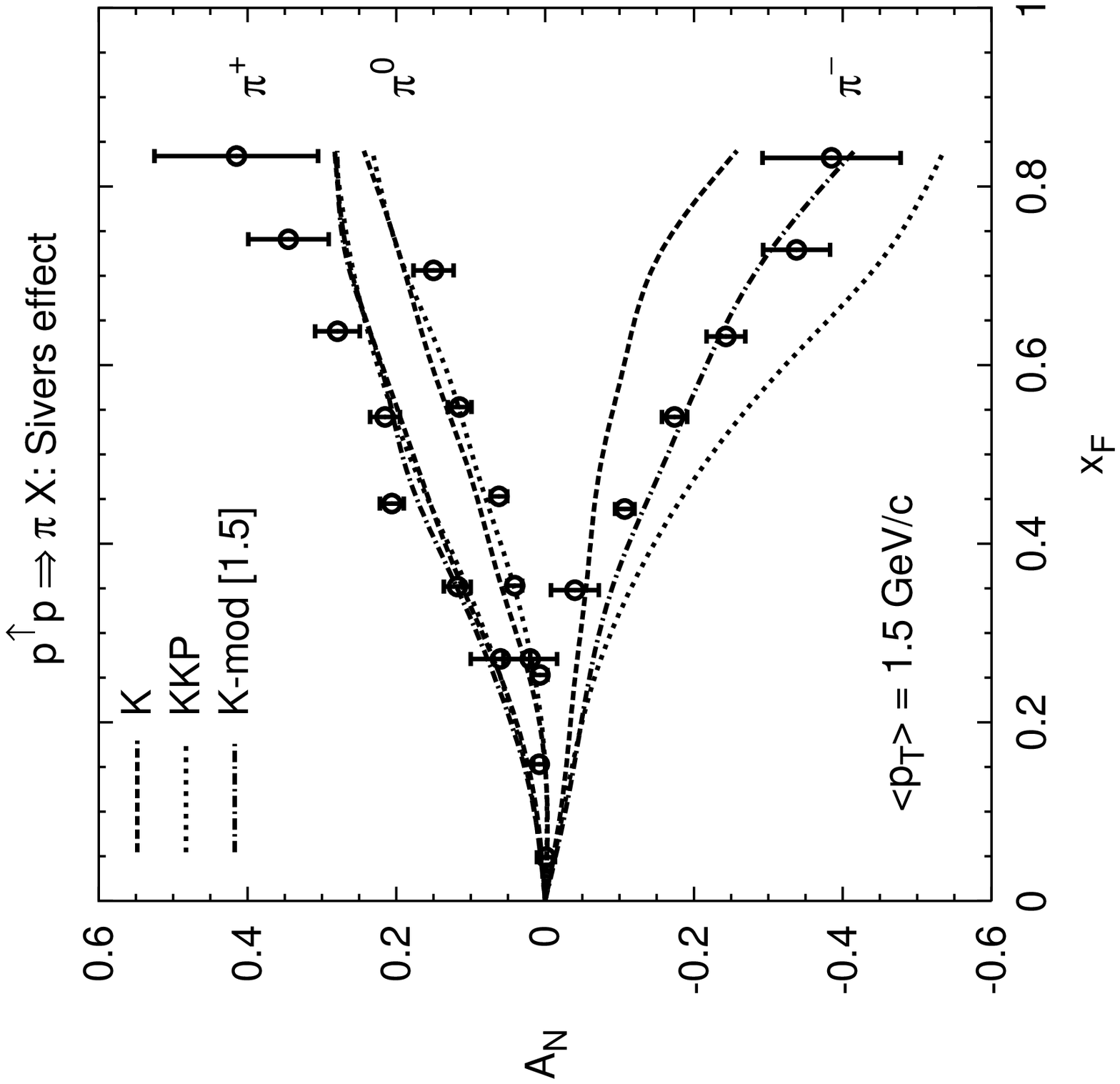}
\includegraphics{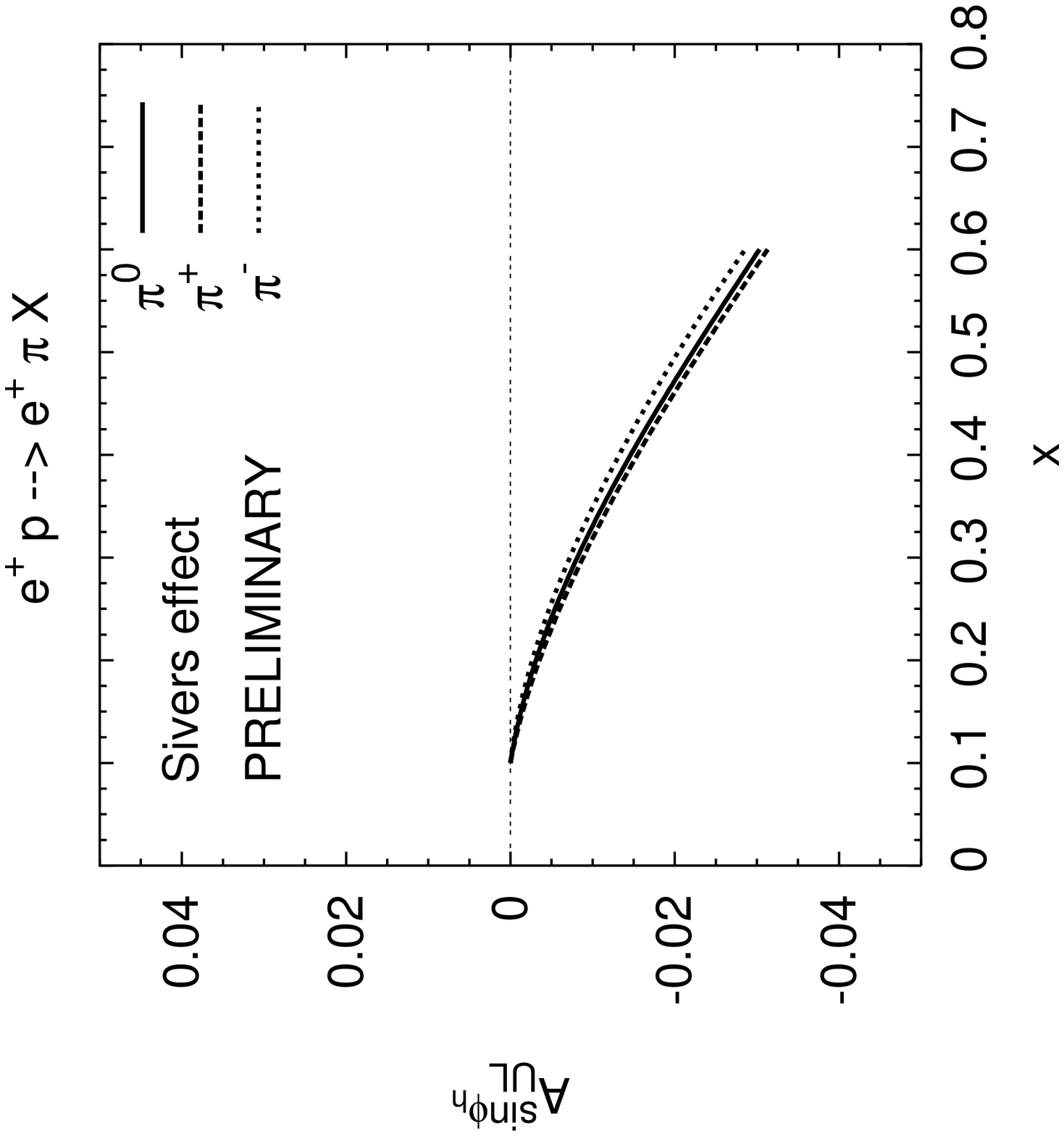}
\caption[*]{ 
SSA in (a) $p^\uparrow\,p\to\pi \,X$ vs. $x_{_F}$ and (b) 
$\ell\,p^\uparrow\to\ell'\pi\,X$ vs. $x$; see plots and text for
more details.}
\end{center}
\vspace*{-0.6cm}
\end{figure}

For the process $p\,p\to\pi\,X$, some experimental results 
for SSA are also available, and we
can see how our approach works for SSA and unpolarized cross sections
at the same time. In this case we can have  
$\bfk_\perp$ effects in the fragmentation process also. 
A direct $z$ dependence of the $\beta'$ parameter in Eq.~(\ref{gk}) 
seems to be favored,
$1/\beta'(z)=\langle\,k_{\perp \pi}^{\,2}(z)\,\rangle^{1/2}=
1.4\,z^{1.3}\,(1-z)^{0.2}$ GeV/$c$.

Let us now consider the SSA in $p^\uparrow\,p\to\pi\,X$, within the same
approach and assuming it is generated by the Sivers 
effect \cite{siv} alone, 
that is from a spin and $\bfk_\perp$ effect in the PDF
inside the initial polarized proton, described by the Sivers function
$\Delta^{\!N}\!f_{q/p^\uparrow}(x,\bfk_\perp)$. Other possible
sources for SSA, like the so-called Collins effect \cite{col}, 
concerning the fragmentation of a polarized parton into the final
observed pion,
are not considered here.
Analogous studies have already been performed \cite{abm}; 
here we show
the first results with full treatment of $\bfk_\perp$ effects
and partonic kinematics. These results are in good qualitative
agreement with previous work. 

The numerator of the SSA, $d\sigma^\uparrow-d\sigma^\downarrow$ can be
expressed in the form of Eq.~(\ref{ucr}), with the substitution
$\hat f_{a/A}(x,\bfk_\perp)\,\to\,
\Delta^{\!N}\!f_{q/p^\uparrow}(x,\bfk_\perp)$. 
For the Sivers function we choose an expression similar to that of the
unpolarized distribution:
\begin{equation}
\Delta^{\!N}\!f_{q/p^\uparrow}(x,\bfk_\perp) \propto
\Delta^{\!N}\!f_{q/p^\uparrow}(x)\,k_\perp\,
\exp\left[\,-\beta^2 k_\perp^2/r\,\right]\,\sin\phi_{k_\perp}\>,
\label{dfh}
\end{equation}
\noindent 
where $\phi_{k_\perp}\!$ is the angle between $\bfk_\perp$ and the
polarization vector of the proton. 
By comparison  with data we can fix both $r$ and  
$\Delta^{\!N}\!f_{q/p^\uparrow}(x)$. 

In Fig. 1a we show our preliminary estimates of $A_N$ with Sivers effect
at $E$ = 200 GeV and $p_{_T}$= 1.5 GeV/c, vs. $x_{_F}$, 
for three different choices of the pion FF:
K \cite{kre}, KKP \cite{kkp} 
and a modified version of K. Data are from \cite{e704}.
The SSA for $\pi^+$ and $\pi^0$ is well reproduced independently of the
FF set. 
Interestingly, the $\pi^-$ case shows a stronger
sensitivity to the relation between the leading and non-leading contributions
to the fragmentation process, which cannot be extracted from present
experimental information. 

A more direct way to extract the Sivers asymmetry 
is the analysis of SSA in 
Drell-Yan processes, that is the production of 
$\ell^+\ell^-$ pairs in the collision of two hadrons $A$ 
and $B$ \cite{adm02}.     
By considering differential cross sections 
in the squared invariant mass ($ M^2 = (p_a + p_b)^2 $), 
rapidity ($ y $) and  
transverse momentum of the lepton pair ($ \bfq_{_T} $) 
and integrating out the di-lepton angular dependence, 
the SSA reads 
($q_{_T}^2 \ll M^2 \ll M_Z^2$ and $k_{\perp a,b}^2 \simeq q_{_T}^2$) 
\be
A_N = \frac
{\sum_q e_q^2 \int d^2\bfk_{\perp q} \>
\Delta^Nf_{q/\Aup}(x_q, \bfk_{\perp q}) \>
\hat f_{\bar q/B}(x_{\bar q}, \bfq_{_T} -\bfk_{\perp q})}
{2 \sum_q e_q^2 \int d^2\bfk_{\perp q} \,
\hat f_{q/A}(x_q, \bfk_{\perp q}) \>
\hat f_{\bar q/B}(x_{\bar q}, \bfq_{_T}-\bfk_{\perp q})} \>. \label{ann}
\ee
Notice that other sources of SSA 
would lead to a contribution to $A_N$ 
that vanishes upon integration 
over all final angular configurations of the $\ell^+\ell^-$ pair.

One further uncertainty concerns the sign of the asymmetry: as noticed 
by Collins and Brodsky \cite{newsiv}, the Sivers asymmetry 
has opposite signs in Drell-Yan and SIDIS, respectively related 
to $s$-channel and $t$-channel elementary reactions. As in $p\,p\to
\pi X$ we expect that large $x_{_F}$ regions are dominated by 
$t$-channel quark processes, we tentatively assume that the Sivers function
extracted from $p-p$ data should be opposite to that contributing
to D-Y processes. 

Finally we consider the Sivers contribution to 
the azimuthal asymmetry in SIDIS.  In this case the numerator of the
transverse SSA reads
\be
 d\Delta\sigma = |\bfS_{_{\!\perp}}|\,
\sum_q e_q^2 \, \Delta^N\! f_{q/p^\uparrow}
(x_q,\bfk_{_{\!\perp}}) \, 
\otimes \frac{d\hat \sigma^{\ell q\to \ell q'}}{d\hat t} \, 
\otimes D_{\pi/q}(z_q,\bfk_{_{\!\perp}}') \label{sidis}\,,
\ee
where $\bfS_\perp$ is the transverse component of the target spin 
in the $\gamma^*-p$ cm frame. In particular for longitudinal
polarization ($\bfS_L$) in the LAB frame 
\be
|\bfS_{_{\perp}}| = |\bfS_L|\,\sin\theta_\gamma
\simeq |\bfS_L|\, 2\,(m_p/Q)\, x\sqrt{1-y}\>.
\label{ST}
\ee 
From Eq.~(\ref{ST}) we see that the contribution in Eq.~(\ref{sidis}) 
is globally higher twist.\footnote{Other possible contributions 
of the same twist are not considered here.}
In Fig 1.b we give our estimate of the ($\sin\phi$)-weighted asymmetry 
$A_{UL}$, for HERMES
kinematics from Eq.~(\ref{sidis}). 
It is important to remark (see also \cite{efr}) 
the tiny value of $A_{UL}$, 
its negative sign
and the $u$ flavor dominance also for $\pi^-$; this last feature is due
to the relatively small difference between favored and 
unfavored FF's in the $z$ range covered at HERMES. 

In conclusion, we have presented here preliminary results of 
a consistent study of partonic transverse momentum effects
both in unpolarized and polarized cross sections (and SSA) for inclusive
particle production in hadronic collisions. 

\section*{Acknowledgments} 
This contribution is based on work done in collaboration with M. Anselmino.
We are also grateful to M. Boglione and E. Leader for useful
discussions. 
Support from COFIN. MURST-PRIN is acknowledged.

\end{document}